\documentstyle[eqsecnum,aps]{revtex}

\begin{document}
\twocolumn
\title{Reversal of thermopower oscillations in mesoscopic Andreev interferometer}
\author{A. Parsons, I. A. Sosnin, and V. T. Petrashov}
\address{
Physics Department, Royal Holloway, University of London, Egham,
Surrey, TW20 0EX, U.K.}
\date{\today}
\maketitle
\begin{abstract}
\par We report measurements of thermopower oscillations $vs$ magnetic field
in a diffusive Andreev interferometer. Upon the increase of the dc
current applied to the heater electrodes, the amplitude of these
oscillations first increases then goes to zero as one would
expect. Surprisingly, the oscillations reappear at yet higher
heater currents with their phase being $\pi$-shifted compared to
low current values. From direct measurements of the temperature
gradient we estimate the amplitude of the oscillations to be
orders of magnitude smaller than predicted by theory.

\end{abstract}


\narrowtext

\par In a nonuniformly heated conductor there arises an electric
field, $E$, proportional to the temperature gradient $E=Q\nabla
T$, where $Q$ is known as thermopower. In metals $Q$ is determined
\cite{1} by a derivative of the logarithm of conductivity $\sigma$
with respect to energy $\varepsilon$ taken at the Fermi level
\begin{equation}
Q=\frac{\pi^{2}}{3}\frac{k_{B}^{2}T}{e}\left\{\frac{\partial
ln\sigma}{\partial
\varepsilon}\right\}_{\varepsilon=\varepsilon_{F}}, \eqnum{1}
\end{equation}
where $k_{B}$ is Boltzman constant and $e$ is electron charge. In
normal metals with diffusive electron transport the conductivity
changes very little with energy and the thermopower has the
following order of magnitude
\begin{equation}
Q=C\frac{k_{B}}{e}\frac{k_{B}T}{\varepsilon_{F}}, \eqnum{2}
\end{equation}
where $C$ is a constant of the order of unity depending on the
topology of Fermi surface and the energy dependence of scattering
time.
\par The thermoelectric properties of a normal
metal ($N$) in contact with a superconductor ($S$) are strongly
modified by the proximity effect. First, the electrical
conductivity has a much stronger energy dependence, so that the
thermopower can be orders of magnitude larger than predicted by
Eq. (2) \cite{2}. In the geometry of Andreev interferometer, when
the normal part is connected to the superconducting loop, the
thermopower will oscillate as a function of the magnetic flux
$\Phi$ through the loop, with a period equal to the flux quantum
$\Phi_{0}=hc/2e$. It was shown that these oscillations can be
symmetric or antisymmetric with respect to $\Phi$ depending on the
sample topology in contrast to the conductance oscillations which
are always symmetric \cite{2}. Second, the voltage between $N$ and
$S$ circuits may appear due to nonequilibrium branch imbalance in
the $N$ film created by temperature gradient \cite{3}. The
thermopower associated with this effect is predicted to be giant
compared with Eq. (2) as it does not contain small factor
$k_{B}T/\varepsilon_{F}$. The thermopower oscillations are
predicted to be close to antisymmetric in this case \cite{3}.
\par Recently, the oscillating thermovoltage
of mesoscopic (Au/Al) Andreev interferometer has been discovered
in a pioneering experiment by Chandrasekhar's group \cite{4}. The
value of $Q$ was estimated to be consistent with theoretical
predictions \cite{2}. For various geometries of Andreev
interferometer, both symmetric and antisymmetric oscillations were
observed. Later experiments by the same group with direct
measurements of temperature gradients proved that the thermopower
was indeed orders of magnitude larger than (2) \cite{5}. The
origin of the phase of thermopower oscillations for different
geometries is still unclear.
\par In this Letter we report measurements of thermopower
oscillations $vs$ magnetic field in a (Sb/Al) Andreev
interferometer. As a function of heater current the amplitude of
oscillations first increases then goes to zero similar to that in
Ref. 4. However, we have discovered a novel effect: at higher
heater currents the oscillations reappear with their phase shifted
by $\pi$ compared to low current ones. In our case the amplitude
of oscillations was of the order of (2), as extrapolated from the
value of $Q$ for Sb at $T=273K$, although classical thermopower
was not observed.
\par The structures were made by multi-layer electron-beam
lithography as shown in the scanning electron micrograph (Fig. 1).
The first layer was 40 nm thick Sb (semimetal) followed by second
layer of 60nm thick Al (superconductor). Prior to the deposition
of the second layer, $in-situ$ Ar$^{+}$ etching was used to clean
the interface. Two hybrid loops form two Andreev interferometers
which we will call "top interferometer" (TI) with interfaces to
superconductor situated on the current lines of $N$-part and
"bottom interferometer" (BI) with the interfaces being off current
lines. TI has $S$-contacts ($S1$ and $S2$) and $N$-contacts ($N1$,
$N2$, $N3$, $N4$, $H1$, $H2$). BI has $S$-contact $S3$ and
$N$-contacts $N5$, $N6$, $N7$, $N8$, $H1$, $H2$ (see Fig. 1). The
geometry of the sample allowed us to measure the temperature
gradient across an interferometer, so that the absolute value of
thermopower could be determined.
\par Measurements were performed in a
He$^{3}$ cryostat in temperatures from 0.28K to 6K with a magnetic
field (< 5T) applied perpendicular to the substrate. Resistivity,
$\rho$, of Sb film was 60$\mu\Omega$cm and that of Al film was
1.2$\mu\Omega$cm, with diffusion constants, $D$, 133cm$^2$/s and
223cm$^2$/s, respectively. The resistance of interface between the
two films in normal state was 8$\Omega$ for the interface area
150x150nm$^{2}$.
\par Figure 2 shows the resistance and thermovoltage oscillations
for both interferometers as a function of the magnetic field, with
a period corresponding to the flux quantum through the
 superconducting loop. Magnetoresistance measurements were performed
  using conventional ac bridge technique. For thermopower measurements
  a heating current, $I_{h}$, was a sum of dc and small
ac currents. Thermovoltage, $V_{th}$, was measured using lock-in
amplifier on the frequency of ac signal.
\par The polarity of the connection of $S$ and $N$ electrodes to the
voltmeter was the same for both TI and BI. Yet, the phase of
thermovoltage oscillations is opposite for TI and BI. If we assume
some heat escape through the $NS$ interfaces into superconductor,
then for BI the closest to the $N$ reservoir $NS$ contact will
have higher temperature, contrary to TI. Thus, the temperature
gradient will be opposite for BI and TI, resulting in opposite
phase of thermopower oscillations for BI and TI. However, for
quasiparticle energies below superconducting gap there should be
no heat transfer into the $S$ contact (neglecting phonon heat
conductivity at $T\approx0.28K$). In this case, there is no
temperature gradient between the $NS$ contacts for BI and the
reason for the opposite phase for BI and TI remains unclear. In
our experiment the oscillations of the thermopower for the BI with
$NS$ interfaces off classical current lines between the $N$
reservoirs (corresponding to the house structure of Ref. 4) were
$\pi /2$-shifted from magnetoresistance oscillations (as opposed
to the two being in phase in Ref. 4). This fact can probably be
attributed to the noticeable asymmetry in the position of $NS$
contacts of BI with respect to the "hot" point (see Fig. 1).
\par Figure 3 shows $V_{th}$ $vs$ magnetic field oscillations of
the TI for four different dc current, $I_{h}$. With increasing
$I_{h}$ the oscillations first disappear, and then remarkably
reappear at a higher $I_{h}$ with a $\pi$ phase shift compared to
low $I_{h}$ measurements. The phase of the thermopower
oscillations at each $I_{h}$ was checked against that of the
magnetoresistance, which remained the same for all temperatures
and currents. Note magnetic field independent $V_{th}$ resulting
in a vertical shift of the curves in Fig. 3 with increasing
$I_{h}$. This is not due to classical thermopower in Sb as control
Sb structures of exactly the same geometry of the heater but with
all normal electrodes (no superconductors) showed no such a shift.
In Fig. 4 we show the amplitude of thermovoltage oscillations for
both our interferometers. Note that the amplitude of the
thermovoltage oscillations for BI and TI is about the same. For
both interferometers no noticeable oscillations were detected for
$I_{h}$ in the range from 3 to 8 $\mu A$.
\par We have also performed similar measurements
of the thermopower on the structures with extra normal electrodes
connected to the Andreev interferometer. This allows us to compare
the thermovoltage arising between two normal electrodes with that
arising between a normal and a superconducting electrode.We found
the amplitude of the thermovoltage oscillations being
approximately the same for both cases, contrary to the prediction
of Ref. {3}. The disagreement probably originates from the
condition $h/\varepsilon \tau _{\varepsilon} \ll 1$ (here $h$ is
Plank's constant, $\varepsilon$ is a characteristic energy of
quasiparticles and $\tau_{\varepsilon}$ is the energy relaxation
time) at which the results of Ref. {3} were calculated \cite{6}.
In our experiment it was $h/\varepsilon \tau _{\varepsilon} \sim
1$.
\par To estimate the absolute value of thermopower we need to know
temperature gradient across the interferometer. We have used
proximity effect in the TI as a thermometer \cite{7}. Figure 5
shows the amplitude of the magnetoresistance oscillations of the
TI as a function of temperature and dc current, $I_{h}$. From this
we can roughly estimate the temperature, $T_{m}$, in the middle of
the normal part of the TI (Fig. 6, inset). Figure 6 shows the
correspondence of the temperature to the heating current extracted
from Fig. 5. The solid line in Fig. 6 shows the best fit of
$T_{m}$ by the formula $\sqrt{T_{0}^{2}+\alpha I_{h}^{2}}$,
obtained from a solution of Nagaev's equation \cite{8} neglecting
electron-phonon scattering at low temperatures, where $\alpha$ is
a sample-specific constant, which we used as a fitting parameter.
We will use the solid line on Fig. 6 to obtain $T_{m}$ at a given
heating current. Broken lines on Fig. 6 show error in finding
$T_{m}$ due to data scattering. The thermopower of Andreev
interferometer, $Q_{A}$ can be estimated as $Q_{A}=V_{th}/\Delta
T$, where $V_{th}$ is voltage measured between $S$ and $N$ and
$\Delta T\approx T_{m}-T_{0}$ is temperature difference across the
interferometer. Measurements on the $N$6-$N$7 part of the
structure confirmed that the temperature of this part does not
deviate from the base temperature at low ($I_{h}\leq 1\mu A$)
heater currents.
\par The reliable estimation of
the temperature using this method can be done only at small
currents when the corresponding temperature is far away from the
critical temperature of the superconducting transition. This is
because close to the superconducting transition the temperature
dependence of the proximity effect is governed by the temperature
dependence of the gap rather than actual electron temperature
\cite{9}. For $I_{h}=1\mu A$ we have $T_{m}\approx 0.36\pm 0.02K$,
so that $\Delta T \approx 80mK$. This gives the value of
$Q_{A}\approx 50nV/K$. It is interesting to compare this value
with the classical thermopower of Sb, $Q_{cl}$. Using table value
of $Q_{cl}=36 \mu V/K$ for Sb at $T=273K$ \cite{10} we can expect
the value of the order of $36 nV/K$ at $T=0.28K$. Thus, in our
experiment the ratio $\varepsilon_{F}/k_{B}T$ seems to be orders
of magnitude larger than $Q_{A}/Q_{cl}$. However, the fact that we
did not observe classical thermovoltage down to the level of about
0.1$nV$, which corresponds to the thermopower of about 1$nV/K$ at
$I_{h}=1\mu A$, suggests that the classical thermopower is also at
least two orders of magnitude smaller than one would expect from
free electron model. Unfortunately, the thermopower measurements
at this temperatures are very difficult, because one needs a small
temperature gradient, resulting in a small thermovoltage. The
reference data for Sb thermopower at these low temperatures is
also lacking.
\par At low heater currents our results are in
general in line with earlier experiment of Ref. \cite{4}. In terms
of thermovoltage our results are of the same order as in \cite{4}
but in terms of thermopower we find our values of 50 - 100nV/K to
be smaller than 4$\mu$V/K reported in \cite{4} and than
theoretical prediction of few $\mu$V/K \cite{2,3}.
\par Main discovery made in this work is reappearance of thermopower
at higher heating currents with the $\pi$-shift in the phase of
the thermopower oscillations. We emphasize that our result is
different from the reversal of Josephson current observed in Ref.
11, because we don't see any anomalies in magnetoresistance
oscillations (see Fig. 5) with their phase being exactly the same
throughout the whole range of temperatures and currents.
\par In real metals with anisotropic Fermi surface and scattering
times (2) will be no longer valid. Instead, we must add up
contributions from all different parts of the Fermi surface, some
having opposite sign. Sb is a highly anisotropic semimetal with
the concentration of electrons and holes being nearly equal and
with effective masses differing by a factor of 10 for different
directions. Contributions to the proximity-effect correction to
conductivity from these different types of carriers will have
different energy dependence. Therefore, it is possible that at
some temperature these contributions may cancel each other and the
thermopower will change sign in the way observed on the
experiment.
\par The other mechanism that strongly affects the thermopower of
metals at low temperatures and may also result in giant
thermopower and a change in the sign of thermopower is a phonon
drag \cite{12}. However, at the base temperature of our
experiment, $T=0.28K$, phonon effects should be minimal. To our
knowledge the phonon drag has never been studied with regard to
superconducting proximity effect.
\par In conclusion, we have observed the reversal of the
phase-dependent thermopower of diffusive Andreev interferometer at
low temperatures. The magnetic-field independent Andreev
thermopower was observed, while classical thermopower was smaller
than the experimental noise level. The amplitude of both Andreev
and classical thermopower was orders of magnitude smaller than
that predicted by theory. We believe that the full theoretical
treatment of the problem including the topology of real Fermi
surface is needed for complete understanding of the observed
effects.
\par We acknowledge financial support from the EPSRC (Grant Ref:
GR/L94611). We thank A.F. Volkov for valuable discussions.

\label{}

\begin{figure}  
\caption{Scanning electron micrograph of the top (TI) and bottom
(BI) interferometers.} \label{Fig. 1.}
\end{figure}

\begin{figure}
\caption{Magnetoresistance and thermovoltage oscillations at
$T=0.28K$. Top panel: TI magnetoresistance measured using current
leads $N1$-$H1$ and potential leads $N4$-$H2$ (solid line); TI
thermovoltage measured using current $H1$-$H2$, potentials
$S2$-$N1$ (broken line). Bottom panel: BI magnetoresistance,
current $N5$-$H1$, potentials $N8$-$H2$ (solid); BI thermovoltage,
current $H1$-$H2$, potentials $S3$-$N5$ (broken).} \label{Fig. 2.}
\end{figure}

\begin{figure}[h]
\caption{Thermovoltage of the TI as a function of magnetic field
for heater currents $I_{h}=1, 5, 12, and 13 \mu A$. $T=0.28K$.}
\label{Fig. 3.}
\end{figure}

\begin{figure}
\caption{Amplitude of thermovoltage oscillations as a function of
dc heater current. Filled circles: TI; Open circles: BI.
$T=0.28K$.} \label{Fig. 4.}
\end{figure}

\begin{figure}
\caption{Reduced amplitude of magnetoresistance oscillations
measured using current leads $N1$-$S1$ and potential leads
$N4$-$S2$ as a function of temperature (left) and heater current
($H1$-$H2$)(right). Modulation ac current was $I_{mod}=0.75\mu A$}
\label{Fig. 5.}
\end{figure}

\begin{figure}
\caption{Circles: Temperature - heater current correspondence
extracted from Fig. 5. Solid line: best fit of $T_{m}$ using
$\alpha=0.049(K/\mu A)^{2}$. Broken lines show inaccuracy in
$T_{m}$ due to data scattering corresponding to
$\alpha=0.070(K/\mu A)^{2}$ and $\alpha=0.033(K/\mu A)^{2}$.}
\label{Fig. 6.}
\end{figure}


\begin{references}

\bibitem{1}N.F. Mott and H. Jones, {\it The Theory of the Properties of Metals
and Alloys} (Clarendon, Oxford, 1936), 1st ed.
\bibitem{2}N.R. Claughton and C.J. Lambert, Phys. Rev. B {\bf53}, 6605 (1996).
\bibitem{3}R. Seviour and A.F. Volkov, Phys. Rev. B {\bf62}, 6116
(2000).
\bibitem{4}J. Eom, C.-J. Chien, and V. Chandrasekhar, Phys. Rev. Lett. {\bf81}, 437 (1998).
\bibitem{5}D.A. Dikin, S. Jung, and V. Chandrasekhar,
cond-mat/0107605 Preprint, 2001.
\bibitem{6}A.F. Volkov, private communications.
\bibitem{7}J. Aumentado, J. Eom, V. Chandrasekhar, P.M. Baldo, and L.E. Rehn, Appl. Phys.
Lett. {\bf75}, 3554 (1999).
\bibitem{8} K.E. Nagaev, Phys. Rev. B {\bf52} 4740 (1995).
\bibitem{9} V.T. Petrashov, R.Sh. Shaikhaidarov, I.A. Sosnin, P.
Delsing, T. Claeson, A. Volkov, Phys. Rev. B {\bf58}, 15088
(1998).
\bibitem{10} Handbook of Chemistry and Physics, Ed. C.D. Hodgman,
Chemical Rubber Publishing Co., Cleveland, Ohio, 1953.
\bibitem{11} J.J.A. Baselmans, A.F. Morpurgo, B.J. van Wees, and T.M.
Klapwijk, Nature {\bf397} 43 (1999).
\bibitem{12}N.A. Red'ko and S.S. Shalyt, Sov. Phys.-Solid State {\bf10}, 1233 (1968).
\end{references}
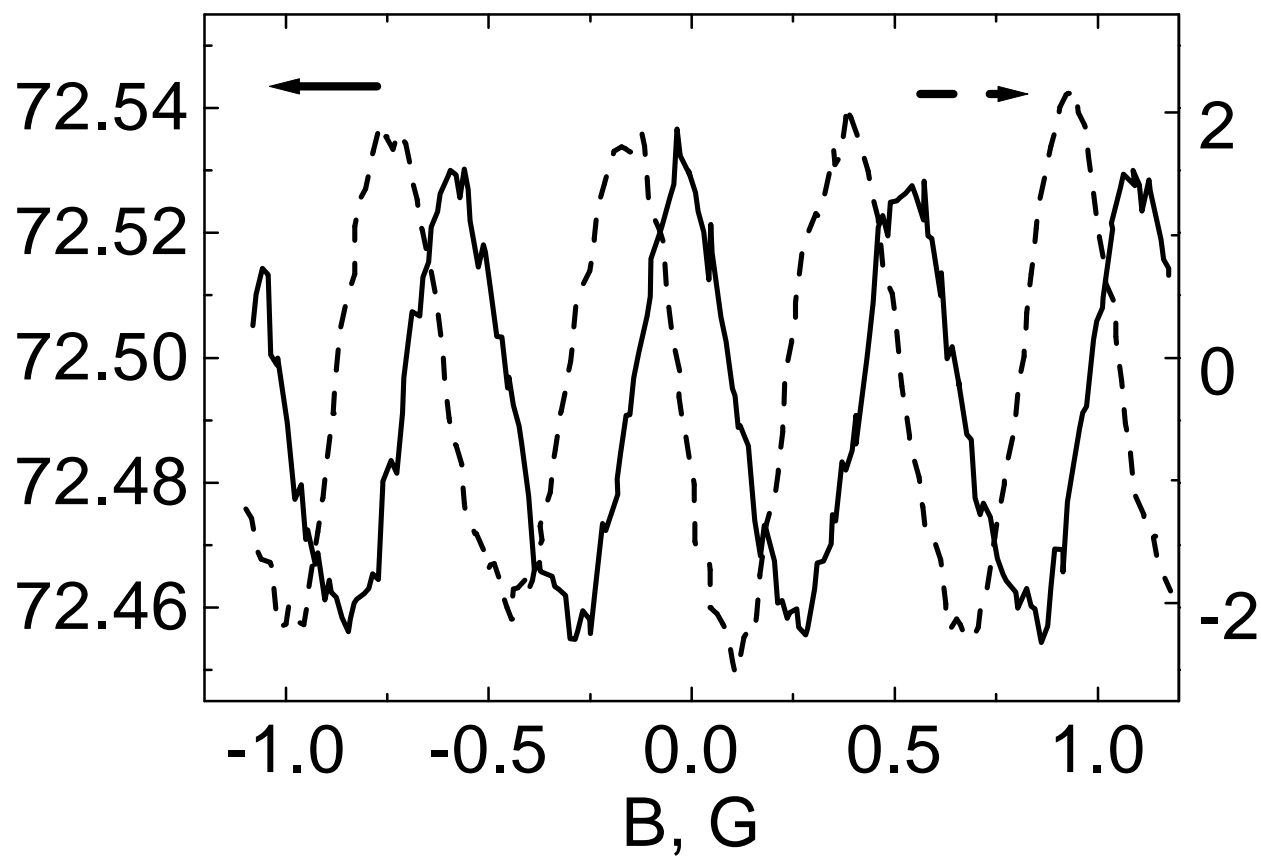
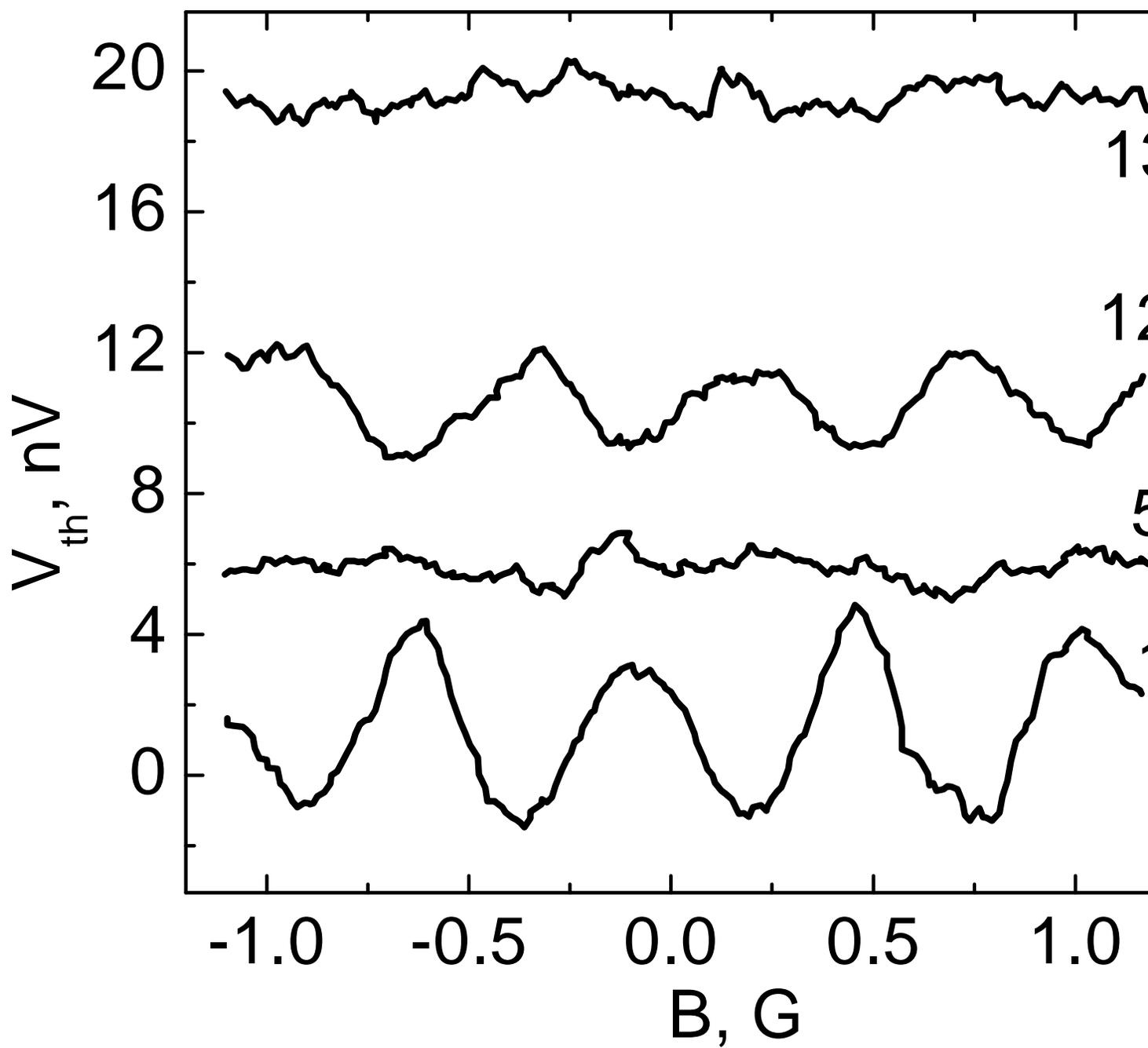
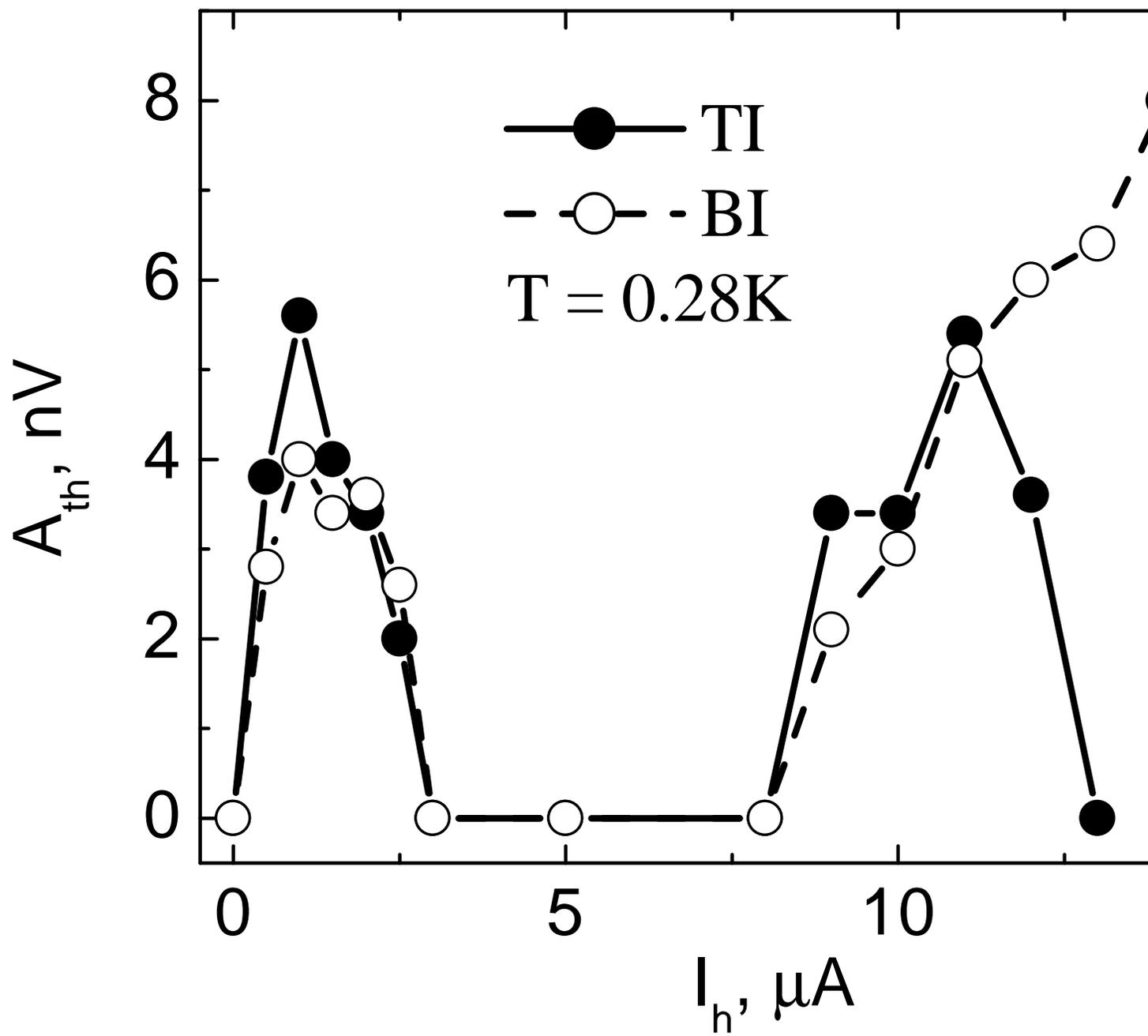
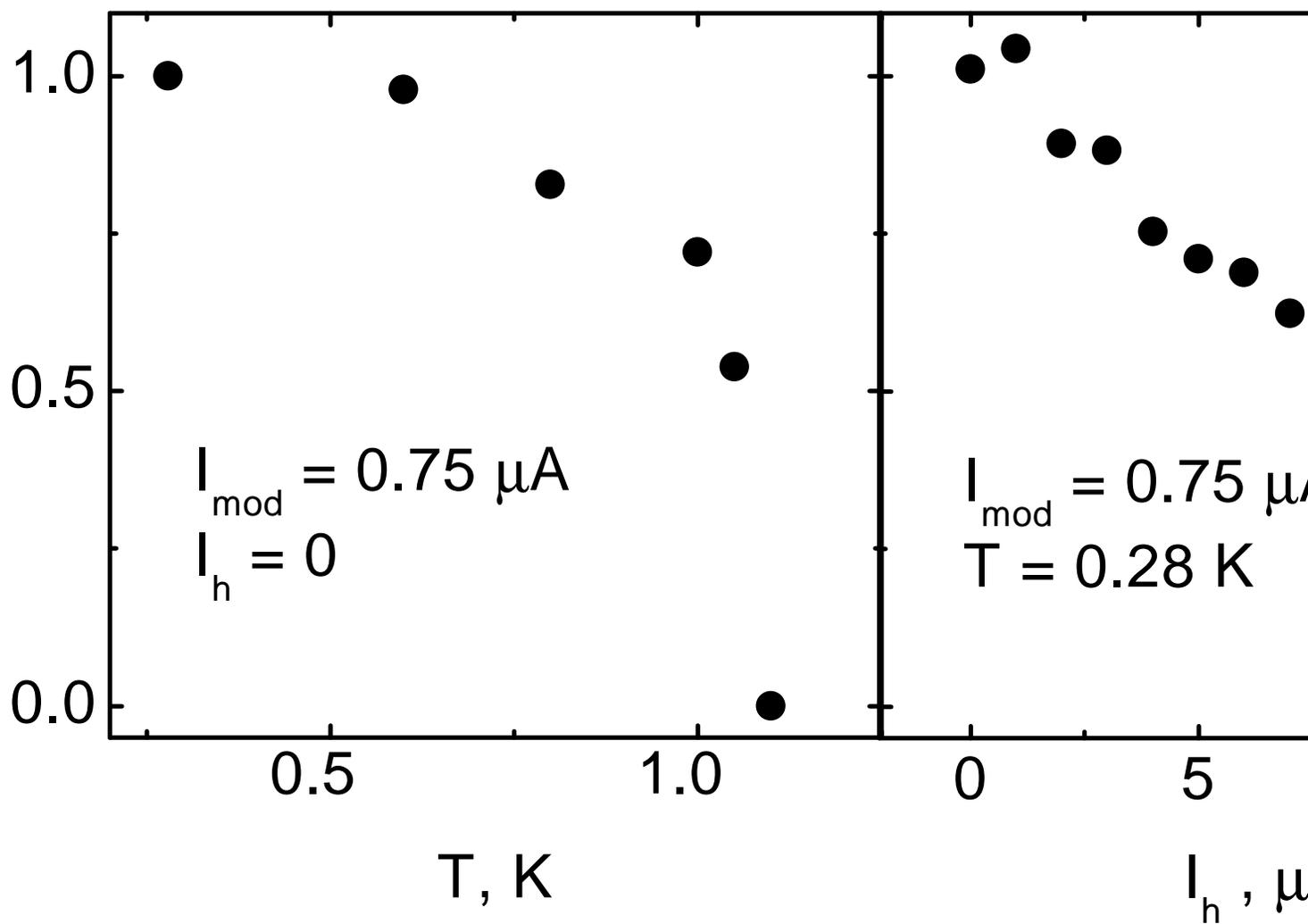
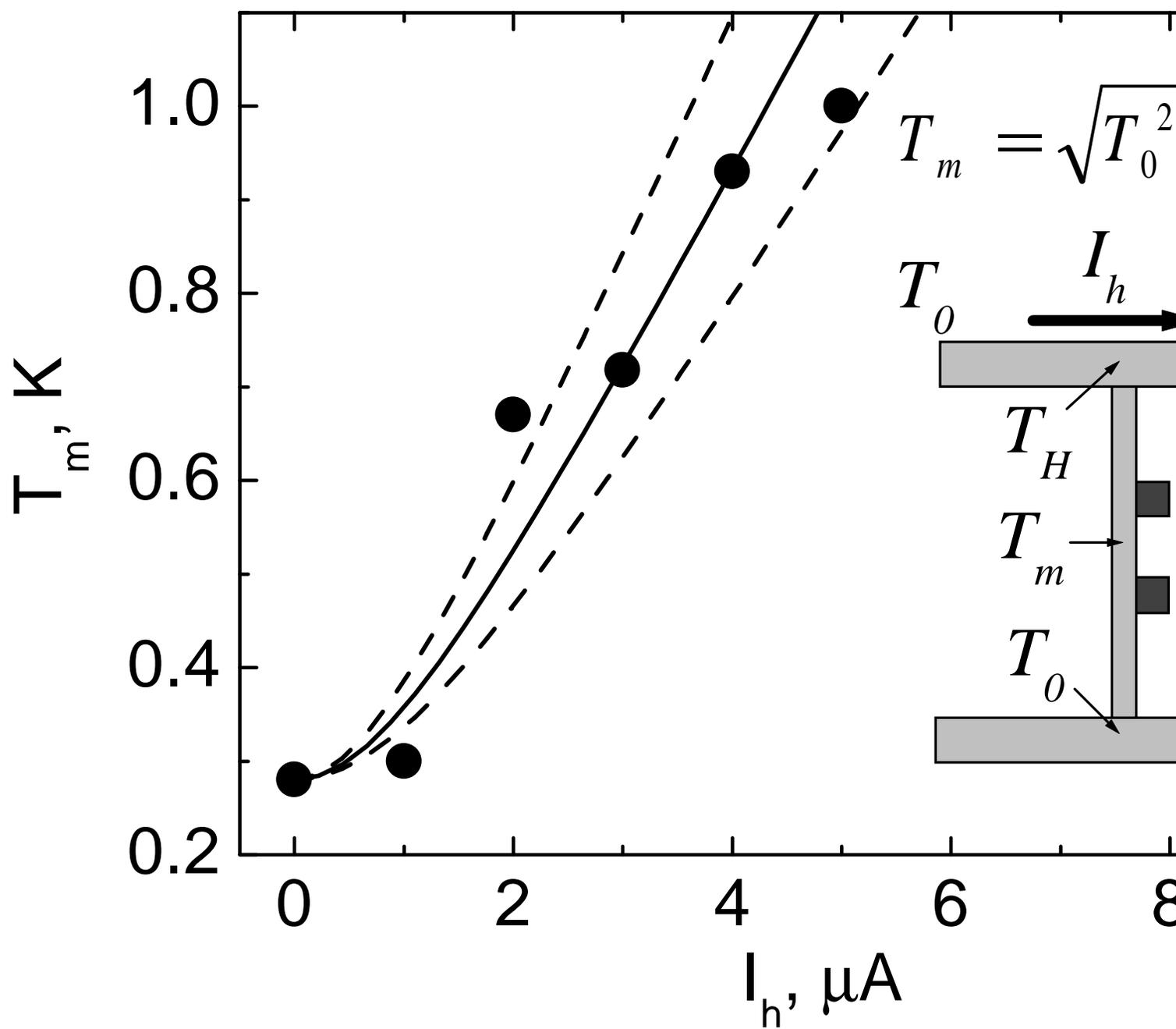
\end{document}